\documentclass[prd,twocolumn,nofootinbib]{revtex4}
\usepackage{graphicx, epsfig}
\usepackage{color}
\usepackage{mathrsfs}
\usepackage{bm}
\usepackage{amsmath,amssymb,empheq}
\usepackage{mathrsfs}
\usepackage[caption=false]{subfig}
\usepackage{hyperref}
\usepackage[normalem]{ulem}
\usepackage{mathtools}
\usepackage[x11names]{xcolor}

\definecolor{fashionfuchsia}{rgb}{0.96, 0.0, 0.63}
\colorlet{no_so_fashion_purple}{blue!50!red}

\newcommand{\be}{\begin{equation}}
\newcommand{\ee}{\end{equation}}
\newcommand{\ba}{\begin{eqnarray}}
\newcommand{\ea}{\end{eqnarray}}

\newcommand{\nn}{\nonumber}

\newcommand{\bmpsi}{{\bm \psi}}
\newcommand{\tildeD}{{\tilde D}}

\newcommand{\barxi}{{\bar \xi}}
\newcommand{\barzeta}{{\bar \zeta}}

\begin{document}
\title{Perturbative stability of non-Abelian electric field solutions}
\author{Jude Pereira\footnote{jperei10@asu.edu}, Tanmay Vachaspati\footnote{tvachasp@asu.edu}}
\affiliation{
Physics Department, Arizona State University, Tempe,  Arizona 85287, USA.
}

\begin{abstract}
We consider SU(2) gauge theory with a scalar field in the fundamental representation. The model
is known to contain electric field solutions sourced by the scalar field that are distinct from
embedded Maxwell electric fields. We examine the perturbative stability of the solution and identify 
a region of parameter space where the solution is stable. 
In the regime where the scalar field has a negative mass squared, the solution has two branches
 and we identify an instability in one of the branches.
\end{abstract}

\maketitle

There is considerable interest in understanding the structure of non-Abelian gauge
theories as these apply to the strong and weak interactions. One aspect of such theories
is the existence of non-trivial classical solutions that might serve as backgrounds for
other quantum phenomena. A uniform electric field in non-Abelian gauge theory 
is one of the simplest such backgrounds but is also sufficiently rich to lead to 
interesting physics. The reason is that in non-Abelian theories there exist multiple
gauge-inequivalent potentials that lead to identical electric fields~\cite{Brown:1979bv}, 
whereas in Abelian
theory there is a unique gauge potential modulo gauge transformations. One can indeed
embed the gauge potential of the Abelian theory into the non-Abelian theory to obtain
an electric field but a separate (infinite) class of gauge potentials also obtain the
same electric field. It is this class of gauge potentials that is of interest to us in this
paper.

Unlike the embedded Abelian gauge potential, the new class of gauge potentials do 
not satisfy the source-free non-Abelian equations of motion\footnote{Even a uniform
electric field in Abelian theory can be viewed as sourced by charges located at infinity. However, the
sources for the electric field derived from the new class of gauge potentials are
space filling.}. These sources may arise as effective degrees of freedom due to
quantum effects or they may be postulated in terms of other fields in the non-Abelian
theory~\cite{Vachaspati:2022ktr}.
Our work examines a recent solution in SU(2) gauge theory with a scalar field transforming 
in the fundamental representation, wherein the scalar field acts as a source for a spatially 
uniform electric field which is derived from gauge potentials in the new 
class~\cite{Vachaspati:2022gco,Vachaspati:2023tpt}. The solution not only solves the 
gauge field equations but also the scalar field equations.


In earlier work~\cite{Pereira:2022lbl}, we had considered the stability of the new class of
gauge potentials and found several instabilities. However, in that work we had restricted
attention to only the gauge fields as then the solution with the fundamental scalar field
was not known. The primary goal of this paper is to examine the stability of the solution 
of the gauge field plus the scalar field. 

The SU(2) theory and the solution contain several parameters such as the gauge coupling
$g$, the scalar mass $m$, the scalar self-coupling $\lambda$, a characteristic frequency
of the solution $\Omega$, and three other parameters of the solution.
In this work, we are able to analyze stability in certain regions of this 
large parameter space. In these regions we find that the solution is stable if
$m^2 \ge 0$. If $m^2 < 0$, instabilities exist for a certain range of values of the other parameters.



\subsection{Electric field solution}
\label{solution}

Consider the Lagrangian for SU(2) gauge theory with a minimally coupled scalar field
in the fundamental representation
\be
L = |D_\mu \Phi |^2 - \frac{1}{4} W_{\mu\nu}^a W^{a \mu\nu} - m^2 |\Phi |^2 - \lambda |\Phi |^4
\ee
where $\Phi$ transforms in the fundamental representation of SU(2)
and $W_\mu^a$ is the SU(2) gauge field that
was analyzed in Ref.~\cite{Vachaspati:2022ktr}
where a solution consisting of a uniform electric field was 
discovered.
The solution is
\be
\Phi = \eta \begin{pmatrix} z_1 e^{+i\omega t} \\ z_2 e^{-i\omega t} 
\end{pmatrix},
\ \ \eta \equiv \frac{\sqrt{2}\, \Omega}{g} 
\label{summaryPhi}
\ee
\be
{\vec W}_\mu = -\frac{\epsilon}{g}  
\left (\cos( \Omega t ),\sin( \Omega t ),0 \right )\partial_\mu z ,
\ \ \epsilon \equiv \sqrt{2\omega\Omega}
\label{summaryW}
\ee
where $z_1,\, z_2 \in \mathbb{C}$ are arbitrary constants subject to the constraint
$|z_1|^2+|z_2|^2=1$, $g$ is the gauge coupling constant that appears in the
covariant derivative $D_\mu = \partial_\mu -ig W^a_\mu \sigma^a/2$, $\sigma^a$
are the Pauli spin matrices, and $\omega$ 
is given in terms of $\Omega$ and the parameters in the scalar potential by 
\be
\omega = \frac{1}{2} \left [ \frac{\Omega}{2} \pm 
\sqrt{\frac{\Omega^2}{4} + 4 \left (m^2 + \frac{4\lambda}{g^2} \Omega^2 \right )} \, \right ] .
\label{omsol}
\ee

The field strength is found using,
\be
W_{\mu\nu}^a = \partial_\mu W_\nu^a - \partial_\nu W_\mu^a 
+ g \epsilon^{abc} W_\mu^b W_\nu^c
\label{fsdefn}
\ee
and yields
\ba
W^1_{\mu\nu} &=& \frac{\sqrt{2\omega\Omega}}{g} 
\Omega \sin(\Omega t) (\partial_\mu t \, \partial_\nu z - \partial_\nu t \, \partial_\mu z ) 
\label{Wmunu1} \\
W^2_{\mu\nu} &=& - \frac{\sqrt{2\omega\Omega}}{g}
\Omega \cos(\Omega t) (\partial_\mu t \, \partial_\nu z - \partial_\nu t \, \partial_\mu z ) 
\label{Wmunu2} \\
W^3_{\mu\nu} &=& 0 
\label{Wmunu3}
\ea
which, by a gauge transformation, becomes~\cite{Vachaspati:2022gco}
\be
W_{\mu\nu}^3 = -E (\partial_\mu t \, \partial_\nu z - \partial_\nu t \, \partial_\mu z )
\ee
and $W_{\mu\nu}^1=0=W_{\mu\nu}^2$, where
\be
E = \frac{\Omega \sqrt{2\omega\Omega}}{g}
\ee
Thus the solution in \eqref{summaryPhi} and \eqref{summaryW} describes 
a uniform electric field in the $z$-direction.

The next question is if the electric field solution is classically stable. This is the
subject of the present analysis. Here we perform a perturbative analysis and
show that there is a range of parameters $(g,m^2,\lambda; \Omega,z_1,z_2)$\footnote{
$g,m^2,\lambda$
are model parameters, whereas $\Omega,z_1,z_2$ are parameters in the solution. 
By rescaling fields and coordinates the number of model parameters can be reduced 
to a single parameter, $\lambda/g^2$, but we have retained them for clarity in taking 
various limits. $z_1$ and $z_2$ define a point on an $S^3$ and can be written in terms 
of three angles and $\Omega$ can take on any real non-negative value.
We will shortly focus on the choice $z_1=1$ and $z_2=0$.} 
for which the solution is classically stable. As long as perturbation theory is valid, the
solution will be quantumly stable as all the perturbative modes will 
correspond to
simple harmonic oscillators with real frequencies. Non-perturbative stability,
{\it e.g.} tunneling to another lower energy state, is a more difficult problem that
we do not address in the present work.

%
%

\subsection{Summary of Results}
\label{summaryresults}

The stability analysis that follows is a highly technical calculation that not
every reader may want to go through. For this reason we summarize our
main results here. We discuss potential relevance of our calculations to 
non-Abelian gauge theories, such as QCD, in Sec.~\ref{conclusions}.

The analysis proceeds by considering perturbations about the background
in \eqref{summaryPhi} and \eqref{summaryW}. Once the equations of
motion and the Gauss constraint equations are expanded and linearized 
in the perturbations, we consider
Fourier modes of the perturbations. The Fourier modes are separated
out into various polarizations, setting up algebraic equations for 13 variables.
Three of these variables, $\alpha_a$, decouple and we can solve a simpler system
of equations that then show that there are no instabilities in this subset
of variables (Sec.~\ref{alpha}). The remaining subset of 10 variables is 
too complicated for a general analysis. However, the system can be
analyzed (Sec.~\ref{gtozero}) in the limit of  weak gauge coupling, $g$,
large scalar mass, $m^2$, and
small parameter $\Omega$. We find that the solution is stable in this
region of parameter space. Another region of parameter space accessible
to analysis is in the long wavelength limit (Sec.~\ref{ktozero}). In this
case, for $m^2 <0$, we find a region of instability that we have plotted in 
Fig.~\ref{instabilityRegion}.

In Sec.~\ref{conclusions} we note that the stability of the solution is likely
due to the effective mass of the gauge bosons due to their interactions
with the scalar field. We also speculate that if there are regions in
parameter space where the uniform electric field is unstable, the
instability might evolve into a configuration where the electric field
forms an Abrikosov lattice~\cite{Abrikosov:1956sx} of electric field flux tubes.
More speculatively, stable electric flux tube 
solutions~\cite{Vachaspati:2022gco,Vachaspati:2023tpt}
may be relevant to confinement in QCD.

\section{Field Perturbations}
\label{perturbations}

We now consider small perturbations around the background,
\be
W_\mu^a = A_\mu^a + q_\mu^a, \ \ \Phi = \Phi_0 + \Psi
\label{WAq}
\ee
where $\{\Phi, A_\mu^a\}$ denotes the solution in \eqref{summaryPhi} and \eqref{summaryW}
and $q_\mu^a$ and $\Psi$ are small perturbations.

The solution in \eqref{summaryPhi} contains the constants $z_1$ and $z_2$, subject only
to the constraint $|z_1|^2+|z_2|^2=1$. A simple choice for these parameters is $z_1=1$
and $z_2=0$. Then
\be
\Phi_0 = \frac{\sqrt{2}\, \Omega}{g} e^{+i\omega t} \begin{pmatrix} 1  \\ 0 \end{pmatrix}
\label{Phi0z20}
\ee
and we write
\be
\Psi = \frac{\sqrt{2}\, \Omega}{g} e^{+i\omega t} \begin{pmatrix} \psi_1+i\psi_2  \\ \psi_3+i\psi_4 \end{pmatrix}
\equiv \frac{\sqrt{2}\, \Omega}{g} e^{+i\omega t} \bmpsi
\ee
where $\psi_i$ ($i=1,\ldots,4$) are real perturbations. The perturbations $\psi_2$, $\psi_3$ and $\psi_4$
are orthogonal to $\Phi_0$ since
\be
\Phi_0^\dag \Psi + \Psi^\dag \Phi_0 = \frac{4\, \Omega^2}{g^2}  \psi_1
\ee
only depends on $\psi_1$. In other words, $\psi_1$ represents perturbations along $\Phi_0$,
while $\psi_2$, $\psi_3$ and $\psi_4$ represent perturbations that are orthogonal to $\Phi_0$.
We will also use the notation
\be
\bmpsi = \begin{pmatrix} \psi_u \\ \psi_d \end{pmatrix}.
\ee

We now turn to the equations of motion for the fields\footnote{We use the mostly minus
signature for the Minkowski metric.}.
The gauge field equation of motion is,
\ba
&& \hskip -0.5 cm
D_\nu W^{\mu\nu a} \equiv \partial_\nu W^{\mu\nu a} + g\epsilon^{abc} W_\nu^b W^{\mu\nu c}
\nn \\ && \hskip 1 cm
=  i \frac{g}{2} \left [ \Phi^\dag \sigma^a D_\mu \Phi - h.c. \right ]
\ea
where $h.c.$ stands for Hermitian conjugate.
The scalar field equations are,
\be
D_\mu D^\mu \Phi + V'(\Phi) =0
\label{Phieom}
\ee
where the prime denotes derivative with respect to $\Phi^\dag$ and
\be
V(\Phi ) = m^2 |\Phi |^2 + \lambda |\Phi |^4.
\label{vphi}
\ee

The equations satisfied by the gauge field perturbations are derived in
Appendix~\ref{appA} and those by the scalar field perturbations in 
Appendix~\ref{appB}.

\section{Equations for the field modes}
\label{modeeqns}

Now that we have the equations for all the perturbations, namely \eqref{Qpmeq},
\eqref{Q3eq}, \eqref{psipmeq} and \eqref{chipmeq}, we expand the perturbations
in plane wave modes. Since the equations are linear, it is sufficient to consider 
the stability of a single plane wave mode. Before implementing the plane wave
expansion, it is helpful to rewrite the temporal ($\mu=0$) and spatial ($\mu=j$)
components separately.

\

$\mu=0$ in \eqref{Qpmeq} gives two Gauss constraint equations:
\ba
&& \hskip -1 cm
\partial_t (\partial_i Q_i^\pm) \pm i \Omega  \partial_i Q_i^\pm  
\pm i\epsilon [ \partial_t Q_z^3 \mp i\Omega Q_z^3 ] \nn \\
&& \hskip 0.75 cm
= 
\mp i\frac{2\Omega^2}{g} (\partial_t \chi + i (\Omega+2\omega) \chi )^\pm 
\label{Qpmeqmueq0}
\ea

$\mu=0$ in \eqref{Q3eq} gives a third Gauss constraint equation:
\ba
&&  \hskip -0.5 cm
\partial_t (\partial_i Q_i^3) +i\frac{\epsilon}{2} \biggl [ 
(\partial_t Q_z^+ + i2\Omega Q_z^+) 
-(\partial_t Q_z^- - i2\Omega Q_z^-) 
\biggr ]
\nn \\ && \hskip 4 cm
= \frac{2\Omega^2}{g} [ \partial_t \psi_2 + 2\omega \psi_1 ] 
\label{Q3eqmueq0}
\ea

$\mu=j$ in \eqref{Qpmeq} gives two propagation equations:
\ba
&&
\square Q_j^{\pm} \pm i 2\Omega \partial_t Q_j^{\pm} + \partial_j (\partial_i Q_i^\pm) 
\nn \\ &&
\pm i\epsilon \biggl [\partial_j Q_z^3 - 2\partial_z Q_j^{3} + \partial_j z \, \partial_i Q_i^3 \nn \\
&& \hskip 0.5 cm
+i\frac{\epsilon}{2} \biggl \{ \partial_j z ( Q_z^+  -Q_z^- )- (Q_j^+ - Q_j^- ) \biggr \} \biggr ] \nn \\
&& \hskip 2 cm
= \mp i\frac{2\Omega^2}{g} \partial_j \chi ^\pm 
+ \frac{2 \epsilon \Omega^2}{g}   \partial_j z \, \psi_1
\label{Qpmeqmueqj}
\ea

$\mu=j$ in \eqref{Q3eq} gives a third propagation equation:
\ba
&&  \hskip -0.5 cm
\square Q_j^{3} + \partial_j (\partial_i Q_i^3) \nn \\
&& \hskip -0.75 cm
+i\frac{\epsilon}{2} \biggl [ \partial_j z \, \partial_i (Q_i^+-Q_i^-) + \partial_j (Q_z^+-Q_z^-) \nn \\
&& \hskip 1 cm
- 2\partial_z (Q_j^{+}-Q_j^{-} ) 
+ i 2\epsilon \partial_j z \, Q_z^3 \biggr ] \nn \\
&& \hskip 1.5 cm
+ (\epsilon^2 + \Omega^2) Q_j^{3}
= \frac{2\Omega^2}{g}  \partial_j \psi_2 
\label{Q3eqmueqj}
\ea

Now we expand the variables in modes with constant coefficients,
\ba
&&
Q_j^a = P_j^a e^{i (\kappa t - k_i x^i )} \nn \\
&&
\psi_1 = \xi_1 e^{i (\kappa t - k_i x^i )} , \ \ 
\psi_2 = \xi_2 e^{i (\kappa t - k_i x^i )} , \nn \\
&&
\chi_1 = \zeta_1 e^{i (\kappa t - k_i x^i )} , \ \ 
\chi_2 = \zeta_2 e^{i (\kappa t - k_i x^i )} ,
\label{modes}
\ea

\begin{figure}
\includegraphics[width=0.30\textwidth,angle=0]{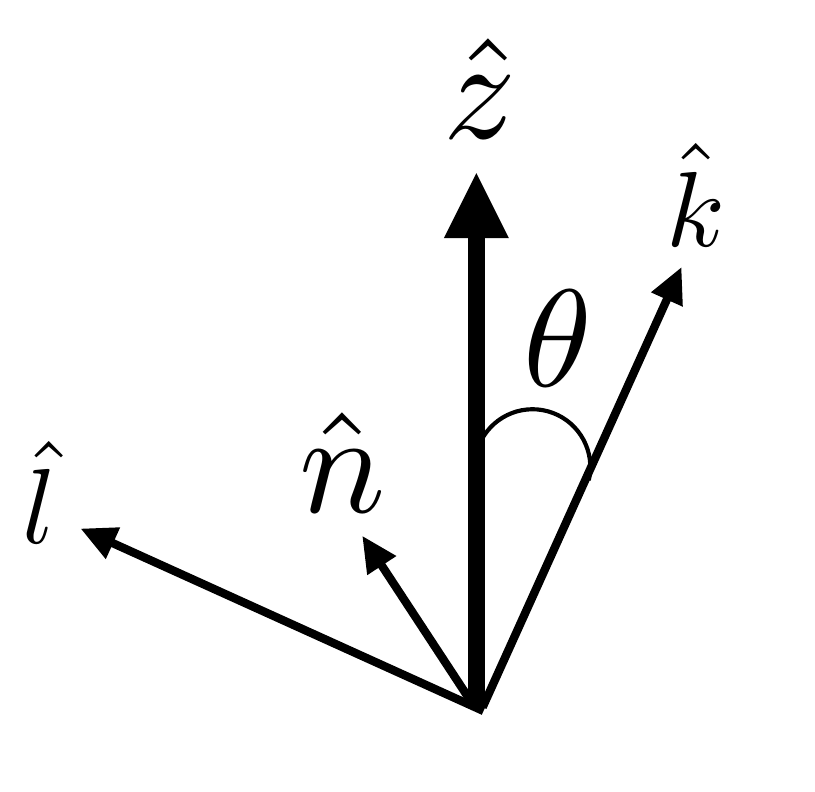}
 \caption{The triad of vectors $({\hat n}, {\hat l}, {\hat k})$. The electric field is along
 the $z$-axis.}
\label{vectorsFigure}
\end{figure}

To separate the various polarizations of the gauge field modes, we define a right-handed 
triad of orthogonal basis vectors $({\hat n}, {\hat l}, {\hat k})$ (See Fig.~\ref{vectorsFigure}),
\be
{\hat n} = {\hat l} \times {\hat k}, \ \
{\hat l} = \frac{{\hat z}- c{\hat k}}{s}, \ \
{\hat k} = \frac{\bf k}{k},  
\ee
where $c \equiv {\hat k}\cdot {\hat z} = \cos\theta$ and $s = \sin\theta$. Then decompose
the Fourier modes in terms of this basis,
\be
P_j^a = \alpha_a {\hat n}_j + \beta_a {\hat l}_j + \gamma_a {\hat k}_j
\ee
and define $\alpha_\pm = \alpha_1 \pm i \alpha_2$ and similarly for $\beta_\pm$
and $\gamma_\pm$. Note that ${\hat z}\cdot {\hat n} =0$.\\
This definition results in the following relations:
\be
k_iP^a_i = k \gamma_a , \quad P^a_z = s \beta_a + c \gamma_a .
\ee

With this decomposition \eqref{Qpmeqmueq0} becomes
\be
(\kappa \pm \Omega  ) k \gamma_\pm
\mp \epsilon ( \kappa \mp \Omega ) (s\beta_3+c\gamma_3)
= \frac{2\Omega^2}{g} [ \kappa  +  \Omega+2\omega ] \zeta_\pm 
\label{gammapmmueq0}
\ee
Eq.~\eqref{Q3eqmueq0} becomes
\ba
&&
\kappa k \gamma_3
- \frac{\epsilon}{2} [ (\kappa + 2 \Omega ) (s\beta_+ +c\gamma_+)
\nn \\ && \hskip 2 cm
-(\kappa - 2 \Omega ) (s\beta_- +c\gamma_-)]
\nn \\ && \hskip 2 cm
= \frac{\Omega^2}{g} [ (\kappa  +  2\omega ) \xi_+ 
- (\kappa  -  2\omega ) \xi_- ]
\label{gamma3eqmueq0}
\ea
Note that $\alpha_a$ do not appear in these $\mu =0$ (constraint) equations.

The ${\hat n}$, ${\hat l}$ and ${\hat k}$ components of Eq.~\eqref{Qpmeqmueqj} become
\be
(-\kappa^2+k^2\mp 2\Omega \kappa ) \alpha_\pm \mp 2\epsilon k_z \alpha_3
\pm \frac{\epsilon^2}{2} (\alpha_+ - \alpha_- )=0.
\label{alphapmeq}
\ee
\ba
&&
(-\kappa^2+k^2\mp 2\Omega \kappa ) \beta_\pm \mp 2\epsilon k_z \beta_3
\pm\epsilon s k \gamma_3 \nn \\ 
&& \hskip -0.5 cm
\pm \frac{\epsilon^2}{2} c [ c (\beta_+-\beta_-) 
-s (\gamma_+-\gamma_-) ] = 2\epsilon \frac{\Omega^2}{g} s \xi_1
\ea
\ba
&&
(-\kappa^2 \mp 2\Omega \kappa ) \gamma_\pm \pm \epsilon s k \beta_3
\nn \\ && \hskip 1 cm
\mp  \frac{\epsilon^2}{2} s [ c (\beta_+ - \beta_-) - s (\gamma_+ - \gamma_-)\} ]
\nn \\ && \hskip 2 cm
 = 2\epsilon \frac{\Omega^2}{g} c \xi_1 - 2 \frac{\Omega^2}{g}  k \zeta_\pm
\ea

Similarly the ${\hat n}$, ${\hat l}$ and ${\hat k}$ components of Eq.~\eqref{Q3eqmueqj} become
\be
(-\kappa^2+k^2 + \epsilon^2 + \Omega^2 ) \alpha_3 - \epsilon k_z (\alpha_+ -\alpha_-)=0.
\label{alpha3eq}
\ee
\ba
&&\hskip -1 cm
(-\kappa^2+k^2 + c^2 \epsilon^2 + \Omega^2 ) \beta_3 
+\frac{\epsilon}{2} \biggl [ s k (\gamma_+ - \gamma_-) 
\nn \\ && \hskip 1.5 cm
- 2 c k (\beta_+ - \beta_-)
- 2 \epsilon s c \gamma_3 \biggr ] =0.
\ea
\ba
&&
(-\kappa^2+ s^2 \epsilon^2 + \Omega^2 ) \gamma_3
+\frac{\epsilon}{2} \biggl [ s k (\beta_+ - \beta_-) 
- 2 \epsilon s c\beta_3 \biggr ] 
\nn \\ && \hskip 4 cm
= -i \frac{2\Omega^2}{g} k \xi_2
\ea

Now for the scalar field modes.
Inserting \eqref{modes} in \eqref{psiueq2} and \eqref{chieq2} 
and writing the equations in terms of $\pm$ variables we get
\ba
&& \hskip -1 cm
(-\kappa^2+ k^2 \mp 2\omega \kappa + 2\lambda \eta^2 )\xi_\pm  + 2\lambda \eta^2  \xi_\mp 
\mp \epsilon c k \zeta_\pm 
\nn \\ && \hskip -1 cm
- \frac{\epsilon g}{4} [ (s \beta_+ +c \gamma_+) + (s \beta_- + c \gamma_-) ] 
\pm \frac{g}{2} k \gamma_3 = 0
\ea

\ba
&& \hskip -1 cm
[-\kappa^2+ k^2  -\Omega (\Omega + 2\omega )  \mp 2(\omega +\Omega) \kappa ] \zeta_\pm
\nn \\ && \hskip 2 cm
\mp \epsilon c k \xi_\pm \pm  \frac{g}{2} k \gamma_\pm = 0
\ea

These equations don't involve the $\alpha_a$ and so the $\alpha_a$ stability
analysis indeed decouples.

For convenience we summarize all the mode equations in Appendix~\ref{appC}.


\section{Stability analysis for $\{ \alpha_a \}$}
\label{alpha}

The system of equations for $\alpha_a$ are decoupled from the other variables and can be solved
independently. We can write the equations as $M_\alpha {\bm \alpha}=0$ where
${\bm \alpha} = (\alpha_+,\alpha_-,\alpha_3)^T$ and
\begin{widetext}
\be
M_\alpha = \begin{pmatrix}
-\kappa^2-  2\Omega \kappa +k^2  +\epsilon^2/2 & - \epsilon^2/2 & - 2\epsilon k_z  \\
- \epsilon^2/2 & -\kappa^2+  2\Omega \kappa +k^2  +\epsilon^2/2 &  2\epsilon k_z  \\
-2\epsilon k_z & 2\epsilon k_z & 2(-\kappa^2 + k^2 + \epsilon^2 + \Omega^2 )
\end{pmatrix}
\ee
\end{widetext}
We require ${\rm Det}(M_\alpha)=0$ and this leads to a cubic equation in $x\equiv \kappa^2$,
\be
x^3 + A x^2 +B x + C = 0
\label{cubic}
\ee
with 
\be
A = -(3 k^2 + 2 \epsilon^2 + 5\Omega^2) <  0
\ee
\be
B = 3 k^4 + 2k^2 (2\epsilon^2 s^2 + 3 \Omega^2) + 5 \Omega^2 \epsilon^2 + \epsilon^4 +4\Omega^4 > 0
\ee
\be
C = -k^2 [ k^4+ (\Omega^2-2\epsilon^2 c_2)k^2 + \epsilon^2 (\epsilon^2+\Omega^2)]
\label{Cform}
\ee
where $c_2 \equiv \cos (2\theta )$.

\begin{figure}
\includegraphics[width=0.40\textwidth,angle=0]{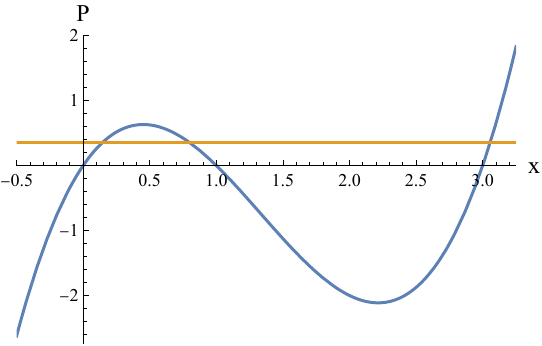}
 \caption{The shape of the cubic curve $P(x)$ in \eqref{Px} (blue curve) and an illustrative
 value of the constant $-C$ in \eqref{Cform} (yellow line). The roots of the cubic equation
 in \eqref{cubic} are given by the intersections of $P(x)$ and the line denoting $-C$. In
 the shown case, all 3 roots of \eqref{cubic} are positive. There would be a negative
 root only if $-C < 0$.}
\label{cubicplot}
\end{figure}

The cubic equation \eqref{cubic} has 3 roots for the $x=\kappa^2$ variable. Two of the
roots may be complex. However, these complex roots are not physical as they don't
satisfy the additional constraint that the gauge fields and scalar field components
are real~\cite{Pereira:2022lbl}. We are only interested in the real roots for $\kappa^2$;
$\kappa$ should be purely real or purely imaginary. If any real root for $\kappa^2$ 
is negative, it will mean that $\kappa = \sqrt{x}$ is imaginary and
that there is an instability. To analyze the roots of the cubic, consider the polynomial
\be
P(x) = x^3 + A x^2 +B x = x (x^2 + A x + B)
\label{Px}
\ee
This polynomial has a root at $x=0$ and the other two roots are given by,
\be
\frac{1}{2} \left [ -A \pm \sqrt{A^2-4 B} \, \right ] \nn
\ee
Since $A< 0$ and $B > 0$, these two roots are either real and positive or they are
complex. Therefore the smallest real root of $P(x)$ is at $x=0$ and the shape
of $P(x)$ is illustrated in Fig.~\ref{cubicplot}. Next, the roots of the cubic in \eqref{cubic} are
given by $P(x) = - C$. Then the cubic has negative real roots only if $-C < 0$,
as shown in Fig.~\ref{cubicplot}. Hence, referring to \eqref{Cform},
 there is an instability if for some values of $k^2 \ge 0$ and 
$-1\le c_2 \le 1$ we can have
\be
k^4+ (\Omega^2-2\epsilon^2 c_2)k^2 + \epsilon^2 (\epsilon^2+\Omega^2) < 0 .
\label{k4}
\ee

The left-hand side of \eqref{k4} is a quadratic in $k^2 > 0$. If 
the discriminant of the quadratic is positive, there will be at least one real and positive
root and the inequality will be satisfied for some $k^2$.
The condition that the discriminant be positive is,
\be
 \frac{1}{4}(2\epsilon^2 c_2-\Omega^2)^2 - \epsilon^2 (\epsilon^2+\Omega^2)  > 0.
\label{condition}
\ee
which implies 
\be
c_2  > \sqrt{1+\frac{\Omega^2}{\epsilon^2}} + \frac{\Omega^2}{2\epsilon^2}
\label{c2>}
\ee
where we have used the condition $2\epsilon^2 c_2 - \Omega^2 > 0$, otherwise
clearly \eqref{k4} cannot be satisfied.
The right-hand side of \eqref{c2>} is greater than 1 but $c_2 = \cos(2\theta ) < 1$
and so there is no instability.

To summarize the results of this section, there are no instabilities in the
$\{\alpha_a \}$ sector of perturbations for any choice of parameters.

\section{Stability analysis for $g, \, \Omega \to 0$}
\label{gtozero}

We first define
\be
\barxi_\pm = \Omega \xi_\pm, \ \ \barzeta_\pm = \Omega \zeta_\pm
\ee
and now $\barxi$ and $\barzeta$ have mass dimension 1.
Next we take the limit,
\be
g \to 0, \ \Omega \to 0, \ E = \frac{\Omega \sqrt{2\omega \Omega}}{g} \ {\rm fixed}
\label{limits}
\ee

We will take $\Omega/g$ fixed in taking the limits in \eqref{limits}. Then to hold $E$ fixed,
we should hold $\epsilon = \sqrt{2\omega \Omega}$ fixed. Therefore
$\omega \sim 1/\Omega \to \infty$. The formula \eqref{omsol} for $\omega$ 
then tells us that either $m^2 \to \infty$ or $\lambda \to \infty$. We choose
to keep $\lambda$ finite and take $m^2 \sim 1/\Omega^2 \to \infty$. 
In this case, $\omega = m \sim 1/\Omega \to \infty$ and $\eta =\sqrt{2}\Omega/g$ is
fixed.

\subsection{Summary of reduced ($g, \, \Omega \to 0$) equations}
\label{summaryg0}

Here we summarize all equations in the limiting case except for the 
$\{ \alpha_a \}$ equations as those have already been dealt with for
general parameters in Sec.~\ref{alpha}.

In taking the limit in the equations we assume that terms such as 
$\Omega g \beta_\pm$, $\Omega g \gamma_\pm$ and $\Omega g \gamma_3$ 
can be neglected because $\Omega g$ is ${\cal O}(g^2)$. This assumption will 
subsequently be checked for consistency.

{\it Constraint equations:}
\be
\kappa k \gamma_\pm \mp \epsilon \kappa (s \beta_3 + c \gamma_3)
= \frac{2\Omega}{g} (\kappa + 2 \omega) \barzeta_\pm
\label{con1g0}
\ee
\ba
&&
\kappa k \gamma_3 - \frac{\epsilon \kappa}{2} \left [ (s\beta_+ + c \gamma_+)
- (s \beta_- + c \gamma_-) \right ] 
\nn \\ && \hskip 1 cm
= \frac{\Omega}{g} \left [ (\kappa + 2\omega ) \barxi_+ - (\kappa - 2\omega ) \barxi_- \right ]
\label{con2g0}
\ea

{\it Equations of motion:}
\ba
&& \hskip -1 cm
(-\kappa^2+k^2 \mp 2\Omega \kappa)\beta_\pm \mp 2\epsilon c k \beta_3 \pm \epsilon s k \gamma_3 
\nn \\ &&
\pm \frac{\epsilon^2 c}{2} \left [ ( c\beta_+ - s\gamma_+ ) - (c \beta_- - s\gamma_-)\right ]
\nn \\ && \hskip 2 cm
- E s (\barxi_+ + \barxi_-) = 0
\label{betapmg0}
\ea
\ba
&& \hskip -1 cm
(-\kappa^2+k^2+\epsilon^2 c^2)\beta_3 
\nn \\ && \hskip -1 cm
+\frac{\epsilon}{2} \left [ s k (\gamma_+-\gamma_-) 
- 2 ck (\beta_+-\beta_-) - 2\epsilon s c \gamma_3 \right ] = 0
\label{beta3g0}
\ea
\ba
&& \hskip -1 cm
(-\kappa^2 \mp 2\Omega\kappa)\gamma_\pm \pm \epsilon s k \beta_3 
\nn \\ &&
\mp \frac{\epsilon^2 s}{2} \left [ (c\beta_+ - s \gamma_+) - (c\beta_- -s \gamma_-) \right ]
\nn \\ && \hskip 2 cm
- E c (\barxi_+ + \barxi_-) + \frac{2\Omega}{g} k \barzeta_\pm = 0
\label{gammapmg0}
\ea
\ba
&& \hskip -1 cm
(-\kappa^2 + \epsilon^2 s^2) \gamma_3
\nn \\ && \hskip -1 cm
+\frac{\epsilon s}{2} \left [ k(\beta_+-\beta_-) - 2\epsilon c \beta_3 \right ]
+ \frac{\Omega k}{g} (\barxi_+-\barxi_-) =0
\label{gamma3g0}
\ea
\be
(-\kappa^2 + k^2 +2\lambda\eta^2-2\omega\kappa) \barxi_+ 
+2\lambda\eta^2 \barxi_- - \epsilon k \barzeta_+ = 0
\label{barxi+}
\ee
\be
(-\kappa^2+k^2+2\lambda\eta^2+2\omega\kappa)\barxi_-
+2\lambda\eta^2\barxi_+ + \epsilon k \barzeta_- = 0
\label{barxi-}
\ee
\be
(-\kappa^2+k^2-\epsilon^2-2\omega\kappa) \barzeta_+ - \epsilon k \barxi_+ =0
\label{barzeta+}
\ee
\be
(-\kappa^2+k^2-\epsilon^2+2\omega \kappa) \barzeta_- + \epsilon k \barxi_- = 0
\label{barzeta-}
\ee

\subsection{Scalar perturbations}
\label{scalarperts}

We see that the $\{ \barxi_\pm, \barzeta_\pm \}$ equations are independent of the
gauge perturbations. Therefore we can solve equations \eqref{barxi+}, \eqref{barxi-},
\eqref{barzeta+} and \eqref{barzeta-} independently of the other equations. And the
$10\times 10$ problem breaks down into a $4\times 4$ problem and a $6\times 6$ problem.
The $4\times 4$ problem is entirely in the scalar sector, {\it i.e.} for the 
$\{ \barxi_\pm, \barzeta_\pm \}$ variables. The $4\times 4$ matrix is,
\begin{widetext}
\be
{\bf M}_\Phi = 
\begin{pmatrix}
-\kappa^2 + k^2 +2\lambda\eta^2-2\omega\kappa & 2\lambda\eta^2 & -\epsilon k & 0 \\
2\lambda\eta^2 & -\kappa^2+k^2+2\lambda\eta^2+2\omega\kappa & 0 & \epsilon k \\
-\epsilon k & 0 & -\kappa^2+k^2-\epsilon^2-2\omega\kappa & 0 \\
0 & \epsilon k & 0 & -\kappa^2+k^2-\epsilon^2+2\omega \kappa
\end{pmatrix}
\label{MPhi}
\ee
\end{widetext}
The trivial solution $\barxi_\pm=0=\barzeta_\pm$ to the $4\times 4$ problem
leads to the $6\times 6$ problem for the variables $\{\beta_a, \gamma_a\}$
which we will deal with in subsection~\ref{gaugepert}.

\begin{figure}
\includegraphics[width=0.40\textwidth,angle=0]{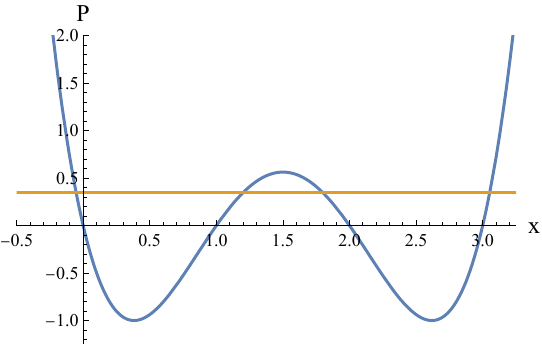}
 \caption{Illustration of the quartic curve $P(x)$ in \eqref{Px} and a horizontal
 line corresponding to $-J$ (defined in \eqref{Jformula}). The intersection
 of $-J$ and the quartic curve give the roots of the quartic in \eqref{quarticpoly}.
 The root for $x < 0$ suggests a possible instability that must be further
 checked to see if it satisfies the Gauss constraints.}
\label{quarticcurve}
\end{figure}

The secular equation is a quartic in $\kappa^2$,
\ba
&& \hskip 0 cm
k^2 [k^2 (k^2 - 2 \epsilon^2)^2 + 4 (k^4 - 3 k^2 \epsilon^2 + 2 \epsilon^4) \eta^2 \lambda ] 
\nn \\ &&
-2 [2 k^6 + 2 \epsilon^4 (\eta^2 \lambda + \omega^2) +   
k^4 (-5 \epsilon^2 + 6 \eta^2 \lambda + 4 \omega^2) 
\nn \\ && \hskip 2.5 cm
+ 2 k^2 (\epsilon^4 - 5 \epsilon^2 \eta^2 \lambda + 4 \eta^2 \lambda \omega^2) ] x
\nn \\ && \hskip -0.5 cm
+[ 6 k^4 + \epsilon^4 -  8 \epsilon^2 (\eta^2 \lambda + \omega^2) +
4 k^2 (-2 \epsilon^2 + 3 \eta^2 \lambda + 4 \omega^2) 
\nn \\ && \hskip 4 cm
+ 16 (\eta^2 \lambda \omega^2 + \omega^4) ] x^2
\nn \\ && 
+[ -4 k^2 + 2 \epsilon^2 - 4 \eta^2 \lambda - 8 \omega^2 ] x^3 + x^4 = 0
\label{quarticpoly}
\ea
where $x \equiv \kappa^2$.
To decide if the system has an instability, the first step is to show that there exists a
negative root of the secular equation. In the $g, \Omega \to 0$ limit, we also have
$\omega \to \infty$, and the secular equation can be written as,
\ba
&&
x[ x^3 - 8 \omega^2 x^2 + 16 \omega^4 x 
- 4 (\epsilon^4 + 2k^4 + 4\lambda \eta^2 k^2 ) \omega^2 ] 
\nn \\ && \hskip -0.5 cm
= - k^2 (k^2-2\epsilon^2) [ k^4 - 2(\epsilon^2 -2\lambda\eta^2) k^2 - 4 \epsilon^2 \lambda \eta^2  ]
\label{quarticanalysis}
\ea
First we only consider the left-hand side of this equation,
\be
P(x) = x[ x^3 - 8 \omega^2 x^2 + 16 \omega^4 x 
- 4 (\epsilon^4 + 2k^4 + 4\lambda \eta^2 k^2 ) \omega^2 ] \nn
\ee
For $\omega \to \infty$, the positions of the extrema of this polynomial are
found by equating the derivative with respect to $x$ to zero. The positions of
the extrema are $x = 2\omega^2, \, 4 \omega^2$ and the extremum at the 
smallest $x$ is at,
\be
x \approx \frac{\epsilon^4 + 2k^4 + 4 \lambda \eta^2 k^2}{8\omega^2} > 0.
\ee
These features imply that the quartic curve given by the left-hand side of 
\eqref{quarticanalysis} has the shape in Fig.~\ref{quarticcurve} -- in particular,
$x=0$ is the smallest root. 

Next we consider the full equation in \eqref{quarticanalysis}. 
Since the smallest root is at very small $|x|$, we can approximate
it by dropping all but the linear term in $x$ on the left hand side of
\eqref{quarticanalysis} to get,
\be
x \approx \frac{k^2 (k^2-2\epsilon^2) [ k^4 - 2(\epsilon^2 -2\lambda\eta^2) k^2 
- 4 \epsilon^2 \lambda \eta^2  ]}
{4 (\epsilon^4 + 2k^4 + 4\lambda \eta^2 k^2 ) \omega^2} .
\label{smallroot}
\ee
The sign of this root will depend on the sign of the numerator (which is
minus the right-hand side of \eqref{quarticanalysis}).
Hence, to find an unstable mode, we need to consider the range of $k^2$, 
$\epsilon^2$ and $2\lambda\eta^2$ for which the numerator of \eqref{smallroot}, 
denoted by $J$, is negative,
\be
J \equiv
k^2 (k^2-2\epsilon^2) (k^2 - k_+^2) (k^2-k_-^2)
\label{Jformula}
\ee
where,
\be
k_\pm^2 = \epsilon^2-2\lambda\eta^2 \pm \sqrt{\epsilon^4+4\lambda^2\eta^4}
\ee
From here we can show that $k_-^2 < 0$ and $k_+^2 > 0$. Therefore the
factor $(k^2-k_-^2)$ in \eqref{Jformula} is positive and the sign of $J$
is the same as the sign of $(k^2-2\epsilon^2)(k^2-k_+^2)$. Some straightforward
algebra shows $k_+^2 < 2 \epsilon^2$. Therefore $J$ is negative for
$k_+^2 < k^2 < 2 \epsilon^2$ and positive otherwise.
Hence there is a possible instability for $k_+^2 < k^2 < 2 \epsilon^2$.

The next step is to find the eigenvectors corresponding to the unstable mode
(negative $\kappa^2$) and check for consistency with the constraint equations.

We note that the possibly unstable mode in \eqref{smallroot} has,
\be
x= \kappa^2 \sim - \frac{1}{\omega^2}
\ee
and we are considering large $\omega^2$. Therefore $-\kappa^2$ is small, 
while $\omega \kappa$ is finite and fixed, even as $\omega \to \infty$.
Then \eqref{barxi+}-\eqref{barzeta-} can be solved to obtain $\barxi_\pm$ 
and $\barzeta_-$ in proportion to $\barzeta_+$. The solutions are,
\be
\barxi_+ = \frac{(k^2-\epsilon^2-2\omega \kappa )}{\epsilon k} \, \barzeta_+ \nn
\ee
\ba
\barxi_- = \frac{\epsilon^2 k^2 - (k^2-2\omega\kappa + 2\lambda \eta^2) 
(k^2-2\omega\kappa -\epsilon^2)}{2\lambda\eta^2 \epsilon k}
\, \barzeta_+ \nn
\ea
\ba
\barzeta_- = \frac{(k^2-2\omega\kappa + 2\lambda \eta^2) (k^2-2\omega\kappa -\epsilon^2)
-\epsilon^2 k^2}{2\lambda\eta^2 (k^2+2\omega\kappa -\epsilon^2)} \,
\barzeta_+ \nn
\ea
The key point here is that $\barxi_\pm$ and $\bar\zeta_-$ are some ${\cal O}(1)$
factors involving $k^2$, $\epsilon^2$, $\omega\kappa$ and $2\lambda\eta^2$
that multiply $\barzeta_+$. In particular, if we set the normalization of the eigenvector
by choosing $\barzeta_+=1$, as we will choose from now on, 
then $\barxi_\pm$ and $\bar\zeta_-$ are also ${\cal O}(1)$.

Next we solve \eqref{betapmg0}-\eqref{gamma3g0}, in the limit that $\kappa \to 0$.
The solution will yield $\beta_a$ and $\gamma_a$ in terms of $\barxi_\pm$ and
$\barzeta_\pm$. From the equations, for $s\ne 0$, we see that $\beta_a$ and 
$\gamma_a$ are the same order as $\barzeta_+$ that we have normalized to 1. 
This justifies the assumption stated  at the beginning of Sec.~\ref{summaryg0}
for $s\ne 0$. For the case when $s=0$, {\it i.e.} for the mode with ${\hat k}={\hat z}$,
Eq.~\eqref{gammapmeq} shows that $\gamma_\pm$ are ${\cal O}(1/g)$. This is
still consistent with taking $\Omega g \gamma_\pm \to 0$ in \eqref{barzetapmeq}
because $\Omega g$ is ${\cal O}(g^2)$ and justifies the assumption even when $s=0$.

The next step is to examine the constraints in Eqs.~\eqref{con1g0}-\eqref{con2g0}.
This immediately leads to a contradiction. For example, the left-hand side of \eqref{con1g0}
goes to zero as $\kappa \to 0$, and forces $\barzeta_\pm \to 0$, which then implies
the trivial case where $\barxi_\pm = 0 = \barzeta_\pm$.

Thus we conclude that there is no instability due to the perturbations $\barxi_a$, $\barzeta_a$,
{\it i.e.} perturbations of the scalar field, and since the equations for these scalar perturbations
do not involve the gauge perturbations, we can set them to zero and separately consider
perturbations of the gauge fields.

\subsection{Gauge field perturbations}
\label{gaugepert}

Now we examine the stability with $\barxi_\pm= 0 =\barzeta_\pm$, that is to perturbations
of the gauge sector only. The Gauss constraints \eqref{con1g0} and \eqref{con2g0} 
now become
\be
k \gamma_\pm \mp \epsilon (s \beta_3 + c \gamma_3) = 0
\label{con1gauge}
\ee
\ba
&&
k \gamma_3 - \frac{\epsilon}{2} \left [ (s\beta_+ + c \gamma_+)
- (s \beta_- + c \gamma_-) \right ]  = 0
\label{con2gauge}
\ea
The equations of motion for $\beta_a$ and $\gamma_a$, given by
\eqref{betapmg0}-\eqref{gamma3g0}, further reduce to
\ba
&& \hskip -1 cm
(-\kappa^2+k^2 \mp 2\Omega \kappa)\beta_\pm \mp 2\epsilon c k \beta_3 \pm \epsilon s k \gamma_3 
\nn \\ &&
\pm \frac{\epsilon^2 c}{2} \left [ ( c\beta_+ - s\gamma_+ ) - (c \beta_- - s\gamma_-)\right ] = 0
\label{betapmg0also}
\ea
\ba
&& \hskip -1 cm
(-\kappa^2+k^2+\epsilon^2 c^2)\beta_3 
\nn \\ && \hskip -1 cm
+\frac{\epsilon}{2} \left [ s k (\gamma_+-\gamma_-) 
- 2 ck (\beta_+-\beta_-) - 2\epsilon s c \gamma_3 \right ] = 0
\label{beta3g0also}
\ea
\ba
&& \hskip -1 cm
(-\kappa^2 \mp 2\Omega\kappa)\gamma_\pm \pm \epsilon s k \beta_3 
\nn \\ &&
\mp \frac{\epsilon^2 s}{2} \left [ (c\beta_+ - s \gamma_+) - (c\beta_- -s \gamma_-) \right ] = 0
\label{gammapmg0also}
\ea
\ba
&& \hskip -1 cm
(-\kappa^2 + \epsilon^2 s^2) \gamma_3
+\frac{\epsilon s}{2} \left [ k(\beta_+-\beta_-) - 2\epsilon c \beta_3 \right ] = 0
\label{gamma3g0also}
\ea

The constraint equations \eqref{con1gauge} and \eqref{con2gauge} are solved
to obtain $\gamma_a$ in terms of $\beta_a$,
\ba
&& \hskip -0.5 cm
\gamma_+ = - \gamma_- = \frac{\epsilon s}{k^2-\epsilon^2 c^2} 
\left [ \epsilon c \frac{(\beta_+ - \beta_-)}{2} + k \beta_3 \right ]
\\ && \hskip -0.5 cm
\gamma_3 = \frac{\epsilon s}{k^2-\epsilon^2 c^2} 
\left [ k \frac{(\beta_+ - \beta_-)}{2} + \epsilon c \beta_3 \right ]
\ea
These relations can now be inserted into Eqs.~\eqref{betapmg0also}-\eqref{gamma3g0also}
to give us 6 equations for 3 variables $\beta_a$. Eqs.~\eqref{betapmg0also}-\eqref{beta3g0also}
lead to
\be
(-\kappa^2 -2\Omega \kappa + k^2+ \epsilon^2/2) \beta_+ - (\epsilon^2/2) \beta_- 
- 2\epsilon c k \beta_3=0
\ee
\be
- (\epsilon^2/2) \beta_+  + (-\kappa^2 + 2\Omega \kappa + k^2+ \epsilon^2/2) \beta_- 
+ 2\epsilon c k \beta_3=0
\ee
\be
- \epsilon c k \beta_+  + \epsilon c k \beta_-  + (-\kappa^2 +  k^2+ \epsilon^2) \beta_3 = 0
\ee
These equations can be written as $M_\beta \beta =0$ where $\beta = (\beta_+,\beta_-,\beta_3)^T$
and
\begin{widetext}
\be
M_\beta = 
\begin{pmatrix}
-\kappa^2 -2\Omega \kappa + k^2+ \epsilon^2/2 & - \epsilon^2/2 & - 2\epsilon c k \\
- \epsilon^2/2 & -\kappa^2 + 2\Omega \kappa + k^2+ \epsilon^2/2 & 2\epsilon c k \\
- 2 \epsilon c k & 2 \epsilon c k & 2 (-\kappa^2 +  k^2+ \epsilon^2)
\end{pmatrix}
\ee
\end{widetext}
Requiring that the determinant of $M_\beta$ vanishes leads to the cubic equation
\ba
&&\hskip -1 cm
x[ x^2 - (3k^2+4\Omega^2+2\epsilon^2) x
\nn \\ && \hskip -0.5 cm
+ (3k^4 +4\epsilon^2 k^2 s^2 + 4\Omega^2 k^2 + 4\Omega^2 \epsilon^2 + \epsilon^4 ) ]
\nn \\ && \hskip 1 cm
= k^2 (k^4-2\epsilon^2 k^2 c_2 + \epsilon^4) 
\label{betacubic}
\ea
where $x \equiv \kappa^2$ and $c_2\equiv \cos(2\theta)$. If the discriminant of the quadratic 
within square brackets on the left-hand side is positive, then both roots are positive, and the
cubic curve has the shape shown in Fig.~\ref{cubicplot} as in the case of the $\alpha_a$
perturbations. The right-hand side of \eqref{betacubic} is easily
shown to be non-negative. Hence the only real root of the cubic equation is positive,
{\it i.e.} $x=\kappa^2 \ge 0$, implying that there is no instability.

\section{Analysis in the $k\to 0$ limit}
\label{ktozero}

The infrared limit can be obtained by taking $k \to 0$ in the mode equations. The resulting equations can be summarized as follows:\\
{\it Constraint equations:}
\be
\hskip -1 cm
\mp \epsilon ( \kappa \mp \Omega ) (s\beta_3+c\gamma_3)
- \frac{2\Omega}{g} [ \kappa  +  \Omega+2\omega ] \barzeta_\pm =0
\label{con1k0}
\ee
\ba
&& \hskip -1 cm
- \frac{\epsilon}{2} [ (\kappa + 2 \Omega ) (s\beta_+ +c\gamma_+)
-(\kappa - 2 \Omega ) (s\beta_- +c\gamma_-)]
\nn \\ && \hskip 1 cm
- \frac{\Omega}{g} [ (\kappa  +  2\omega ) \barxi_+ 
- (\kappa  -  2\omega ) \barxi_- ] = 0
\label{con2k0}
\ea

{\it Equations of motion:}
\ba
&& \hskip -1 cm
(-\kappa^2+k^2\mp 2\Omega \kappa ) \beta_\pm \pm \frac{\epsilon^2}{2} c [ c (\beta_+-\beta_-) 
-s (\gamma_+-\gamma_-) ]  \nn \\ 
&& \hskip 3 cm
- \epsilon \frac{\Omega}{g} s (\barxi_+ + \barxi_-) = 0
\label{betapmk0}
\ea
\ba
&&\hskip 0 cm
(-\kappa^2 + c^2 \epsilon^2 + \Omega^2 ) \beta_3
- \epsilon^2 s c \gamma_3 =0.
\label{beta3k0}
\ea

\ba
&&
(-\kappa^2 \mp 2\Omega \kappa ) \gamma_\pm \mp  \frac{\epsilon^2}{2} s [ c (\beta_+ - \beta_-) - s (\gamma_+ - \gamma_-)\} ]
\nn \\ && \hskip 3 cm
 - \epsilon \frac{\Omega}{g} c (\barxi_+ + \barxi_-) = 0
 \label{gammapmk0}
 \ea
\ba
&&
(-\kappa^2+ s^2 \epsilon^2 + \Omega^2 ) \gamma_3
- \epsilon^2 s c\beta_3 = 0
\label{gamma3k0}
\ea

\ba
&& \hskip -1 cm
(-\kappa^2 \mp 2\omega \kappa + 2\lambda \eta^2 )\barxi_\pm  + 2\lambda \eta^2  \barxi_\mp 
\nn \\ && \hskip 0 cm
- \frac{\epsilon \Omega g}{4} [ (s \beta_+ +c \gamma_+) + (s \beta_- + c \gamma_-) ] = 0
\label{barxipmk0}
\ea

\ba
&& \hskip -1 cm
[-\kappa^2 -\Omega (\Omega + 2\omega )  \mp 2(\omega +\Omega) \kappa ] \barzeta_\pm = 0
\label{barzetapmk0}
\ea
Thus, we see that the complete system of equations has decoupled into three separate sectors $\{\barzeta_{\pm}\}$, $\{\beta_3,\gamma_3\}$ $\{\beta_{\pm},\gamma_{\pm},\barxi_{\pm}\}$. We will analyze each sector separately and look for instabilities.
\subsubsection{$\{\barzeta_{\pm}\}$ sector}
\label{barzetapmsec}

Assuming $\barzeta_{\pm} \neq 0$, we set
\be
\kappa^2 \pm 2(\omega +\Omega) \kappa +\Omega (\Omega + 2\omega ) = 0
\ee
The quadratic equation in $\kappa$ can be solved to obtain $\kappa = \pm \Omega$ or $\kappa=\pm(\Omega + 2\omega)$. Since $\kappa$ is real in both cases, there are no instabilities, and we can proceed by setting $\barzeta_{\pm} = 0$ in what follows.

\subsubsection{$\{\beta_3,\gamma_3\}$ sector}
\label{beta3gamma3sec}
The $2 \times 2$ matrix corresponding to \eqref{beta3k0} and \eqref{gamma3k0} can be written as
\be
M_2 = 
\begin{pmatrix}
-\kappa^2 + c^2 \epsilon^2 + \Omega^2 & - \epsilon^2 s c \\
- \epsilon^2 s c & -\kappa^2+ s^2 \epsilon^2 + \Omega^2
\end{pmatrix}
\label{M2}
\ee
Setting $\text{det}M_3 = 0$ gives
\be
(\kappa^2 - \Omega^2)(-\kappa^2 + \epsilon^2 + \Omega^2) = 0
\ee
which can be solved to obtain $\kappa^2 = \Omega^2$ and $\kappa^2 = \epsilon^2 + \Omega^2$ and hence there are no instabilities.

\subsubsection{$\{\beta_{\pm},\gamma_{\pm},\barxi_{\pm}\}$ sector}
\label{bigsec}
The $6 \times 6$ matrix corresponding to the system of equations \eqref{betapmk0}, \eqref{gammapmk0} and \eqref{barzetapmk0} can be written as
\begin{widetext}
\be
M_6 = 
\begin{pmatrix}
X^{(-)}_c & -\epsilon^2 c^2 / 2 & -\epsilon^2cs / 2 & \epsilon^2cs / 2 & - \epsilon s \Omega / g & - \epsilon s \Omega / g \\
-\epsilon^2 c^2 / 2 & X^{(+)}_c & - \epsilon^2 cs / 2 & - \epsilon\Omega s / g & - \epsilon \Omega s / g \\
-\epsilon^2 sc / 2 & \epsilon^2 sc / 2 & X^{(-)}_s & -\epsilon\Omega c / g & - \epsilon \Omega c / g \\
\epsilon^2 sc / 2 & -\epsilon^2 sc / 2 & -\epsilon^2s^2 / 2 & X^{(+)}_s & - \epsilon\Omega c / g & -\epsilon \Omega c / g \\
-\epsilon s \Omega g / 4 & -\epsilon s \Omega g / 4 & - \epsilon c \Omega g / 4 & - \epsilon c \Omega g / 4 &Y^{(-)} & 2\lambda \eta^2 \\
-\epsilon s \Omega g / 4 & -\epsilon s \Omega g / 4 & - \epsilon c \Omega g / 4 & - \epsilon c \Omega g / 4 & 2\lambda\eta^2 & Y^{(+)}
\end{pmatrix}
\label{M6}
\ee
\end{widetext}
where $X^{(\pm)}_c \equiv -\kappa^2 \pm 2\Omega\kappa + \epsilon^2 c^2 / 2 $,
$X^{(\pm)}_s \equiv -\kappa^2 \pm 2\Omega\kappa + \epsilon^2 s^2 / 2 $, and
$Y^{(\pm)} \equiv  -\kappa^2 \pm 2\omega\kappa + 2\lambda\eta^2$.
Though this matrix might seem complicated at first glance, the secular equation simplifies to give
\be
\kappa^6(\epsilon^2 - \kappa^2 +4\Omega^2)\big[\epsilon^2\Omega^2 - (\kappa^2 - 4\Omega^2)(\kappa^2 - 4\omega^2 - 4\lambda\eta^2)\big] = 0
\ee
Setting the first two terms in the product to zero yields $\kappa^2 = 0$ (trivial solution) and $\kappa^2 = \epsilon^2 + 4\Omega^2$ (stable solution). The third term can be written as
\be
x^2 - 4  (\lambda\eta^2 + \omega^2 + \Omega^2)x + (16\lambda\eta^2 + \omega^2 - \epsilon^2) \Omega^2 = 0
\ee
where $x\equiv \kappa^2$. This quadratic equation in $x$ can be solved to obtain
\be
\kappa_{\pm}^2 = 2\lambda \eta^2 + 2\omega^2 + 2\Omega^2 \pm \sqrt{\epsilon^2\Omega^2 + (2\lambda\eta^2 + 2\omega^2 - 2\Omega^2)^2}
\ee
One can check that $\kappa_{+}^2 > 0$ and hence is stable. 
However, it is possible for $\kappa_-^2 < 0$ provided
\be
16 (\lambda \eta^2 + \omega^2) < \epsilon^2
\ee
and $\Omega \neq 0$. Using $\epsilon^2 = 2\Omega\omega$, we can rewrite the inequality as
\be
(\omega - \omega_{+})(\omega - \omega_{-}) < 0
\label{omquad}
\ee
where we have defined
\be\label{omegapm}
\omega_{\pm} = \frac{\Omega}{16}(1 \pm \sqrt{1 - 512 \lambda / g^2})
\ee
For $\lambda/g^2 < 1/512$ the term in the bracket in 
\eqref{omegapm} will always be positive for real values of 
$\omega_{\pm}$ and hence we have $\omega_{\pm}/\Omega > 0$.

\begin{figure}
\includegraphics[width=0.40\textwidth,angle=0]{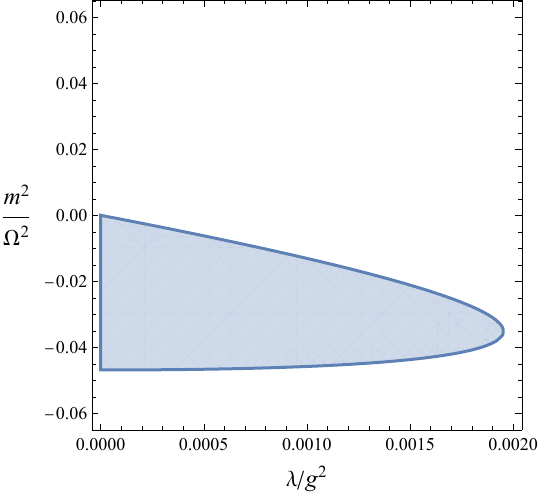}
 \caption{Domain of instability corresponding to the inequality in \eqref{dominst} with minus sign solution for $\omega$. The unstable region is confined to negative values of $m^2$.}
\label{instabilityRegion}
\end{figure}

We can assume $\Omega > 0$ without loss of generality, in which case, the values of $\omega$ that lie in the range $(\omega_{-}, \omega_{+})$ result in instabilities. With the solution for $\omega$ in \eqref{omsol}, the domain of instability is found to be
\be\label{dominst}
1 - \sqrt{1 - 512 a} < 4(1 \pm \sqrt{1+ 64a + 16b}) < 1 + \sqrt{1 - 512 a}
\ee
where $a\equiv \lambda/g^2$ and $b \equiv m^2/\Omega^2$. This region of instability is plotted in Fig~\ref{instabilityRegion}. One can check that, for the allowed values of $a$ and $b$, the plus sign in the solution for $\omega$ in \eqref{omsol} does not give unstable modes.

However, it is not sufficient to just identify the unstable modes. One must also check that the instability persists even after the constraint equation \eqref{con2k0} is imposed. Our approach will be to first solve $M_6V_6 = 0$ after setting $\kappa = \kappa_{-}$, where we have defined $V_6^T = (\beta_{+}, \beta_{-}, \gamma_{+}, \gamma_{-}, \xi_{+}, \xi_{-})$. Then, the solution for $V_6$ will be substituted in \eqref{con2k0} to constrain the allowed values of $c=\cos\theta$, $s=\sin\theta$.  Accordingly, we find the solution
\be
V_6 = \beta_{+}
\begin{pmatrix}
1 \\
(\kappa_{-} + 2\Omega) / (\kappa_{-}-2\Omega) \\
c/s \\
(\kappa_{-} + 2\Omega)c / (\kappa_{-}-2\Omega)s \\
-g(\kappa_{-}-2\omega)(\kappa_{-} + 2\Omega) / 2\epsilon s\Omega \\
-g(\kappa_{-}+2\omega)(\kappa_{-} + 2\Omega) / 2\epsilon s\Omega
\end{pmatrix}
\label{V6}
\ee
One can check that this solution identically satisfies \eqref{con2k0} for all values of 
$c$ (and $s$) and so the instabilities shown in Fig.~\ref{instabilityRegion} satisfy the
Gauss constraints.

\section{Conclusions}
\label{conclusions}
We have examined the classical perturbative stability of the electric field solution
in \eqref{summaryPhi} and \eqref{summaryW}. The general stability problem is technically
difficult but we have shown analytically that there is a range of parameter space, 
namely small coupling constant $g$, large scalar mass 
$m^2$, and small solution parameter $\Omega$, where there is no classical instability. 
One way to understand the stability versus the instability found in Ref.~\cite{Pereira:2022lbl} 
is that the analysis in~\cite{Pereira:2022lbl} took fixed current sources that 
are independent of the gauge field. In contrast,
the currents for the electric field in the present analysis involve the covariant
derivatives of the scalar field which contain the gauge field. Effectively, the
non-vanishing scalar field $\Phi$ gives a mass to the gauge field and suppresses
instabilities. This is apparent in the detailed calculations. For example, the diagonal 
terms in \eqref{MPhi} have $2\lambda\eta^2$ contributions. While this physical
argument suggests
stability, a full stability calculation is necessary together with careful consideration
of the Gauss constraints. Indeed, we found an unstable mode in the dynamical
equations of motion in Sec.~\ref{scalarperts} that turned out to violate the Gauss 
constraints and hence was unphysical.

In Sec.~\ref{ktozero} we have examined the stability of the electric field in the
$k \to 0$ limit without placing any other restrictions on the model or solution parameters.
In this case, there are no instabilities if $m^2 > 0$. There are two branches of electric field
solutions if $m^2 < 0$ given by the $\pm$ signs in \eqref{omsol}. There are
no instabilities if we choose the $+$ sign but there are instabilities for the solution
with the $-$ sign for any direction of the wave vector ${\vec k}$. The parameter
space for which there is an instability is shown in Fig.~\ref{instabilityRegion}.

A stable classical solution provides a background on which quantum effects can
be examined. Usually the classical background solution is chosen to be the trivial 
one, $W_\mu^a=0$, $\Phi=0$, but non-trivial topological defect backgrounds are 
also considered~\cite{Coleman_1985}. Since our electric field solution is time-dependent
but stationary, it has been argued that it is also stable to quantum decay by
Schwinger pair production~\cite{Schwinger:1951nm}
of gauge bosons~\cite{Vachaspati:2022gco}. It is likely that 
Schwinger pair production of scalar particles will be absent or suppressed, at least 
in the corner of parameter space where the scalar mass is 
large\footnote{Electric fields of the Maxwell type are, however,
unstable to rapid Schwinger pair production of gauge particles~\cite{Cardona:2021ovn}.}. 

If there are parameters for which the electric field is unstable, it may point to
an analogy with a uniform magnetic field in a Type II superconductor which is 
unstable to breaking up into an Abrikosov lattice of magnetic 
vortices~\cite{Abrikosov:1956sx}. Perhaps 
there is a range of parameters for which the uniform non-Abelian electric field 
is unstable to breaking up into an Abrikosov lattice of electric vortices of the 
type discussed in Ref.~\cite{Vachaspati:2022gco}. It would be interesting to
map out the stability properties of the electric field over the entire range of
parameter space.

Another direction to investigate is the generalization of our solution
to larger gauge groups such as SU(3) and to examine possible 
relevance to QCD where quark confinement is due to the presence
of electric flux tubes. In QCD there are no fundamental scalar fields but
the fermionic quarks transform in the fundamental representation of 
color SU(3). It is an interesting question if fermionic quark fields
can also provide suitable sources for the new class of color gauge potentials
and color electric fields.

\,

\acknowledgements
TV is grateful to Tufts Institute of Cosmology for hospitality while this work was
being done.
This work was supported by the U.S. Department of Energy, Office of High Energy 
Physics, under Award No.~DE-SC0019470.

\appendix

\section{Gauge field equations}
\label{appA}

The perturbed gauge field equation is
\ba
&& \hskip - 1 cm
 \partial_\nu q^{\mu\nu a} + g \epsilon^{abc} (A_\nu^b q^{\mu\nu c} + q_\nu^b A^{\mu\nu c}) 
 \equiv \delta j^{\mu a} \nn \\
&&=i \frac{g}{2} \left [ \Phi_0^\dag \sigma^a \partial^\mu\Psi + \Psi^\dag \sigma^a \partial^\mu \Phi_0 - h.c. \right ] \nn \\
&& \hskip 0.5 cm
+\frac{g^2}{2} \left [ |\Phi_0|^2 q^{\mu a} + (\Phi_0^\dag\Psi + \Psi^\dag \Phi_0) A^{\mu a} \right ]
\label{gaugeeq}
\ea
where
\be
q_{\mu\nu}^a = \partial_\mu q_\nu^a - \partial_\nu q_\mu^a + g\epsilon^{abc} (A_\mu^b q_\nu^c+q_\mu^b A_\nu^c) 
\ee

It is convenient to work with $q_\mu^{\pm} = q_\mu^1 \pm i q_\mu^2$. Other $\pm$ variables
are defined in a similar way and, for example,
\ba
&& \hskip -0.75 cm
q_{\mu\nu}^\pm = \partial_\mu q_\nu^\pm - \partial_\nu q_\mu^\pm \nn \\
&& \hskip 1 cm
\mp ig [ A_\mu^\pm q_\nu^3 + A_\nu^3 q_\mu^\pm - A_\nu^\pm q_\mu^3 - A_\mu^3 q_\nu^\pm ]
\\ &&  \hskip -0.75 cm
q_{\mu\nu}^3 = \partial_\mu q_\nu^3 - \partial_\nu q_\mu^3  \nn \\
&& \hskip 1 cm
+ i\frac{g}{2} [ A_\mu^+ q_\nu^- + A_\nu^- q_\mu^+ - A_\nu^+ q_\mu^-  - A_\mu^- q_\nu^+  ]
\ea

The gauge equation \eqref{gaugeeq} gives,
\ba
&&
\partial_\nu q^{\mu\nu \pm} \nn \\
&&
\pm ig [ A^{\mu\nu\pm} q_\nu^3 + A_\nu^3 q^{\mu\nu\pm} - A^{\mu\nu 3} q_\nu^\pm - A_\nu^\pm q^{\mu\nu 3} ] 
- \Omega^2 q^{\mu \pm}  \nn \\
&&
= i \frac{g}{2} [ \Phi_0^\dag \sigma^\pm \partial^\mu\Psi + \Psi^\dag\sigma^\pm \partial^\mu \Phi_0
- \partial^\mu\Psi^\dag \sigma^\pm \Phi_0 - \partial^\mu\Phi_0^\dag\sigma^\pm \Psi ] \nn \\
&&
+ \frac{g^2}{2} (\Phi_0^\dag \Psi + \Psi^\dag \Phi_0 ) A^{\mu \pm} 
\label{qpmeq}
\ea
and,
\ba
&& \hskip -0.25 cm
 \partial_\nu q^{\mu\nu 3}  \nn \\
 &&
 + i \frac{g}{2} [ A_\nu^+ q^{\mu\nu -} - A_\nu^- q^{\mu\nu +} + q_\nu^+ A^{\mu\nu -} - q_\nu^- A^{\mu\nu +} ] \nn \\
&&=i \frac{g}{2} \left [ \Phi_0^\dag \sigma^3 \partial^\mu\Psi + \Psi^\dag \sigma^3 \partial^\mu \Phi_0 - h.c. \right ] \nn \\
&& \hskip 0.25 cm
+\frac{g^2}{2} \left [ |\Phi_0|^2 q^{\mu 3} + (\Phi_0^\dag\Psi + \Psi^\dag \Phi_0) A^{\mu 3} \right ] .
\label{q3eq}
\ea

The unperturbed solution for the scalar field is given in \eqref{Phi0z20} and
in temporal gauge the gauge field is
\be
A^\pm_\mu = -\frac{\epsilon}{g}  e^{\pm i \Omega t} \, \partial_\mu z , \ \ 
A^3_\mu =0, 
\ee
\be
A_{\mu\nu}^\pm = \pm i \frac{\epsilon\Omega}{g} e^{\pm i\Omega t} 
(\partial_\mu z \partial_\nu t - \partial_\nu z \partial_\mu t ), \ \ A_{\mu\nu}^3 =0
\ee
The choice of temporal gauge implies $q_0^a=0$. 

Inserting the unperturbed solution in \eqref{qpmeq} gives,
\ba
&&  \hskip -0.5 cm
\square q^{\mu\pm} + \partial^\mu (\partial_i q_i^\pm) 
\mp i \epsilon e^{\pm i\Omega t} (\partial_z q^{\mu 3} - \partial^\mu z \, \partial_i q_i^3 ) \nn \\
&& \hskip 0 cm
\mp i\epsilon e^{\pm i\Omega t} \biggl [\pm i\Omega \partial^\mu t \, q_z^3 - \partial^\mu q_z^3 +\partial_z q^{\mu 3} \nn \\
&& \hskip -0.5 cm
+i\frac{\epsilon}{2} \biggl \{ e^{i\Omega t} \left ( \partial^\mu z \, q_z^- - q^{\mu -} \right )
- e^{- i\Omega t}\left ( \partial^\mu z \, q_z^+ - q^{\mu +} \right ) \biggr \} \biggr ] \nn \\
&& 
+\Omega^2 q^{\mu \pm} = 
\mp i\frac{2\Omega^2}{g} (\partial^\mu \psi_d + i 2\omega \partial^\mu t \, \psi_d )^\pm
\nn \\  && \hskip 2 cm
+ \frac{2 \epsilon \Omega^2}{g}  e^{\pm i\Omega t} \psi_1 \partial^\mu z 
\ea
where
\ba
&& \hskip -0.5 cm
(\partial^\mu \psi_d + i 2\omega \partial^\mu t \, \psi_d )^\pm \equiv
(\partial^\mu \psi_3 -  2\omega \partial^\mu t \, \psi_4 ) \nn \\ 
&& \hskip 3 cm
\pm i (\partial^\mu \psi_4 +  2\omega \partial^\mu t \, \psi_3 )
\ea
It is nicer to extract the $e^{\pm i \Omega t}$ dependence by defining,
\be
q_\mu^\pm = e^{\pm i\Omega t} Q_\mu^\pm, \ \ q_\mu^3 = Q_\mu^3 , \ \ 
\psi_d = e^{i \Omega t} \chi .
\label{chivar}
\ee
Then,
\ba
&&
\square Q^{\mu\pm} \pm i 2\Omega \partial_t Q^{\mu\pm}  \nn \\
&&
+\partial^\mu (\partial_i Q_i^\pm) \pm i \Omega \partial^\mu t \, \partial_i Q_i^\pm 
 + \Omega\epsilon \partial^\mu t \, Q_z^3 \nn \\
&&
\pm i\epsilon \biggl [\partial^\mu Q_z^3 - 2\partial_z Q^{\mu 3} + \partial^\mu z \, \partial_i Q_i^3 \nn \\
&& \hskip 0.5 cm
+i\frac{\epsilon}{2} \biggl \{ \partial^\mu z ( Q_z^+  -Q_z^- )
- (Q^{\mu +} - Q^{\mu -} ) \biggr \} \biggr ] \nn \\
&& \hskip -0.75 cm
= 
\mp i\frac{2\Omega^2}{g} (\partial^\mu \chi + i (\Omega+2\omega) \partial^\mu t \, \chi )^\pm 
+ \frac{2 \epsilon \Omega^2}{g}   \partial^\mu z \, \psi_1
\label{Qpmeq}
\ea
where
\be
(\partial^\mu \chi + i (\Omega+2\omega) \partial^\mu t \, \chi )^+ =
\partial^\mu \chi +i (\Omega+2\omega) \partial^\mu t \, \chi
\ee
\be
(\partial^\mu \chi + i (\Omega+2\omega) \partial^\mu t \, \chi )^- =
\partial^\mu \chi^* -i (\Omega+2\omega) \partial^\mu t \, \chi^*
\ee

Inserting the unperturbed solution and \eqref{chivar} in \eqref{q3eq} gives,
\ba
&&  \hskip -0.5 cm
\square Q^{\mu3} + \partial^\mu (\partial_i Q_i^3) \nn \\
&& \hskip -0.75 cm
+i\frac{\epsilon}{2} \biggl [ \partial^\mu z \, \partial_i (Q_i^+-Q_i^-) + \partial^\mu (Q_z^+-Q_z^-) \nn \\
&& \hskip -0.75 cm
- 2\partial_z (Q^{\mu +}-Q^{\mu -} ) + i2\Omega \partial^\mu t (Q_z^++Q_z^-) 
+ i 2\epsilon \partial^\mu z \, Q_z^3 \biggr ] \nn \\
&&
+ (\epsilon^2 + \Omega^2) Q^{\mu 3}
= \frac{2\Omega^2}{g} [ \partial^\mu \psi_2 + 2\omega \partial^\mu t \, \psi_1 ] 
\label{Q3eq}
\ea

Equations~\eqref{Qpmeq} and \eqref{Q3eq} are our final equations for the
gauge field perturbations.

\section{Scalar field equations}
\label{appB}

The covariant derivative can be written as
\be
D_\mu = \partial_\mu - i \frac{g}{2} W_\mu^a \sigma^a
= \tildeD_\mu - i \frac{g}{2} q_\mu^a \sigma^a
\ee
where
\be
\tildeD_\mu = \partial_\mu - i \frac{g}{2} A_\mu^a \sigma^a.
\ee
Also note
\be
A_\mu^a \sigma^a = -\frac{\epsilon}{2g} ( e^{+i\Omega t}\sigma^- + e^{-i\Omega t}\sigma^+) \partial_\mu z
\ee
\be
q_\mu^a \sigma^a = -\frac{1}{2} ( e^{+i\Omega t}Q_\mu^+ \sigma^- + e^{-i\Omega t}Q_\mu^- \sigma^+)
+ Q_\mu^3 \sigma^3
\ee
and
\be
\sigma^+ = 2\begin{pmatrix} 0&1\\0&0 \end{pmatrix}, \ \ 
\sigma^- = 2\begin{pmatrix} 0&0\\1&0 \end{pmatrix}
\ee

To first order in perturbations,
\ba
D_\mu D^\mu \Phi &-& \tildeD_\mu \tildeD^\mu \Phi_0  =
\nn \\ 
&& \hskip -1.75 cm
+ \tildeD_\mu \tildeD^\mu \Psi 
- i \frac{g}{2} \tildeD_\mu (q^{\mu a}\sigma^a \Phi_0) - i \frac{g}{2} q_\mu^a\sigma^a  \tildeD^\mu \Phi_0
\label{DDPhi}
\ea

The first term on the right-hand side of \eqref{DDPhi} expands to,
\ba
&&
(\tildeD_\mu\tildeD^\mu \Psi)_\uparrow = 
\frac{\sqrt{2}\Omega}{g} e^{i\omega t} 
\biggl [ \square \psi_u + \left ( \frac{\epsilon^2}{4} - \omega^2 \right ) \psi_u 
\nn \\ && \hskip 4 cm
+i 2\omega \partial_t \psi_u -i\epsilon \partial_z \chi \biggr ]
\ea
\ba
&&
(\tildeD_\mu\tildeD^\mu \Psi)_\downarrow = \frac{\sqrt{2}\Omega}{g} e^{i\omega_+ t} 
\biggl [ \square \chi + \left ( \frac{\epsilon^2}{4} - \omega_+^2 \right ) \chi 
\nn \\ && \hskip 3 cm
+i 2\omega_+ \partial_t \chi -i\epsilon \partial_z \psi_u \biggr ]
\ea
where the $\uparrow$ and $\downarrow$ subscripts denote the upper and lower components
of the doublet and we have defined 
\be
\omega_+ \equiv \omega+\Omega .
\ee

The second term on the right-hand side of \eqref{DDPhi} expands to,
\ba
&& \hskip -0.5 cm
-i \frac{g}{2} \tildeD_\mu (q^{\mu a}\sigma^a \Phi_0) =
\nn \\ && \hskip 1.5 cm
i\frac{\Omega}{\sqrt{2}} e^{i\omega t} 
\begin{pmatrix}
\partial_i Q_i^3 + i \epsilon Q_z^+ /2 \\
e^{i\Omega t} ( \partial_i Q_i^+ + i \epsilon Q_z^3 /2 )
\end{pmatrix}
\ea

The third term on the right-hand side of \eqref{DDPhi} expands to
\ba
- i \frac{g}{2} ( q_\mu^a\sigma^a  \tildeD^\mu \Phi_0 )_\uparrow =
- \frac{\epsilon\Omega}{2\sqrt{2}} e^{i\omega t} Q_z^-
\ea
\ba
- i \frac{g}{2} ( q_\mu^a\sigma^a  \tildeD^\mu \Phi_0 )_\downarrow =
\frac{\epsilon \Omega}{2\sqrt{2}} e^{i\omega_+t} Q_z^3
\ea

Putting all the terms together we get for the first order terms,
\ba
&&
\delta (D_\mu D^\mu \Phi)_\uparrow = 
\frac{\sqrt{2}\Omega}{g} e^{i\omega t} 
\biggl [ 
\nn \\ &&
\square \psi_u + \left ( \frac{\epsilon^2}{4} - \omega^2 \right ) \psi_u 
+i 2\omega \partial_t \psi_u -i\epsilon \partial_z \chi 
\nn \\ && \hskip 2 cm
+ i\frac{g}{2} (\partial_i Q_i^3 + i \frac{\epsilon}{2} Q_z^+ )
- \frac{\epsilon g}{4} Q_z^-
\biggr ] 
\ea
\ba
&&
\delta (D_\mu D^\mu \Phi)_\downarrow = 
\frac{\sqrt{2}\Omega}{g} e^{i\omega_+ t} 
\biggl [ 
\nn \\ &&
\square \chi + \left ( \frac{\epsilon^2}{4} - \omega_+^2 \right ) \chi 
+i 2\omega_+ \partial_t \chi -i\epsilon \partial_z \psi_u
\nn \\ && \hskip 5 cm
+ i\frac{g}{2} \partial_i Q_i^+  \biggr ] 
\ea

Finally we find the potential term to first order,
\be
\delta V' = (m^2+2\lambda |\Phi_0|^2)\Psi + 2\lambda (\Phi_0^\dag \Psi + \Psi^\dag \Phi_0)\Phi_0
\ee
or
\ba
&& \hskip -1 cm
\delta V'_\uparrow = \frac{\sqrt{2}\Omega}{g} e^{i\omega t} [ 
( m^2 + 6 \lambda\eta^2 ) \psi_1 
+ i ( m^2 + 2\lambda\eta^2 ) \psi_2 ]
\ea
\be
\delta V'_\downarrow = \frac{\sqrt{2}\Omega}{g} e^{i\omega_+ t} 
(m^2 + 2 \lambda\eta^2 ) \chi
\ee
where we have used $\eta^2= 2\Omega^2/g^2$.
Hence the scalar perturbation equations are,

\ba
&&
\square \psi_u + \left ( \frac{\epsilon^2}{4} - \omega^2 + m^2 + 2\lambda\eta^2 \right ) \psi_u 
+i 2\omega \partial_t \psi_u -i\epsilon \partial_z \chi 
\nn \\ &&
+ i\frac{g}{2}\partial_i Q_i^3   - \frac{\epsilon g}{4} (Q_z^+ + Q_z^-)
+ 4\lambda\eta^2 \psi_1 = 0
\label{psiueq}
\ea
\ba
&& \hskip -0.5 cm
\square \chi + \left ( \frac{\epsilon^2}{4} - \omega_+^2 + m^2 + 2 \lambda\eta^2 \right ) \chi 
+i 2\omega_+ \partial_t \chi
\nn \\ && \hskip 1.5 cm
-i\epsilon \partial_z \psi_u + i\frac{g}{2}\partial_i Q_i^+ = 0
\label{chieq}
\ea

The expression for $\omega$ in \eqref{omsol} follows from the quadratic equation
\be
\frac{\epsilon^2}{4}-\omega^2 +m^2+2\lambda \eta^2 =0.
\ee
where we recall $\epsilon^2 = 2\omega\Omega$. Using this relation,
\ba
&&\hskip -1.5 cm
\square \psi_u  + 4\lambda\eta^2 \psi_1 
+i 2\omega \partial_t \psi_u -i\epsilon \partial_z \chi 
\nn \\ &&\hskip 0.5 cm
+ i\frac{g}{2}\partial_i Q_i^3   - \frac{\epsilon g}{4} (Q_z^+ + Q_z^-)
= 0
\label{psiueq2}
\ea
\ba
&& \hskip -0.5 cm
\square \chi - \Omega (\Omega + 2\omega ) \chi
+i 2\omega_+ \partial_t \chi
-i\epsilon \partial_z \psi_u 
\nn \\ && \hskip 3.5 cm
+ i\frac{g}{2}\partial_i Q_i^+ = 0
\label{chieq2}
\ea

These equations may be written in terms of $\pm$ variables as,
\ba
&& \hskip -1 cm
(\square + 2\lambda \eta^2 ) \psi_\pm \pm i 2\omega \partial_t \psi_\pm
+2\lambda\eta^2 \psi_\mp 
\nn \\ && \hskip 0 cm
\mp i\epsilon \partial_z \chi_\pm 
- \frac{\epsilon g}{4} (Q_z^+ + Q_z^-) \pm i \frac{g}{2} \partial_i Q_i^3 = 0
\label{psipmeq}
\ea
\ba
&& \hskip -1 cm
\square \chi_\pm - \Omega (\Omega+2\omega)\chi_\pm \pm i2\omega_+ \partial_t \chi_\pm
\nn \\ && 
\mp i \epsilon \partial_z \psi_\pm \pm i \frac{g}{2} \partial_i Q_i^\pm = 0
\label{chipmeq}
\ea
where $\psi_\pm=\psi_1\pm i \psi_2$ and $\chi_\pm = \chi_1 \pm i \chi_2$.

This implies that the mass squared in the $\psi_1$ equation is
$4\lambda \eta^2$ while it is zero in the $\psi_2$ equation. In the $\chi$
equation the squared mass is $-\Omega (\Omega + 2\omega) < 0$ suggesting that 
there is an
instability but the equation also has the $+i 2\omega_+ \partial_t \chi$ term
and this makes it less obvious. For example, if the equation was simply
\be
\square \chi - \Omega(\Omega + 2\omega ) \chi +i 2\omega_+ \partial_t \chi  = 0
\ee
and we write $\chi = A \exp(i\kappa t)$, then we get
\be
-\kappa^2 - \Omega (\Omega +2\omega ) - 2 (\omega+\Omega) \kappa =0
\ee
and this has real roots $\kappa = -\Omega, \, -\Omega-2\omega$ and there is no 
instability. In fact, taking $Q^a_\mu=0$, $\psi_u=0$ and $\kappa = -\Omega - 2 \omega$
is a solution to all the equations. This solution corresponds to a perturbation of $\Phi_0$
such that now there is a non-zero $z_2$.

\section{Summary of mode equations}
\label{appC}

We first define
\be
\barxi_\pm = \Omega \xi_\pm, \ \ \barzeta_\pm = \Omega \zeta_\pm
\ee
and now $\barxi$ and $\barzeta$ have mass dimension 1. 

{\it $\alpha_a$ equations:}
\be
(-\kappa^2+k^2\mp 2\Omega \kappa ) \alpha_\pm \mp 2\epsilon k_z \alpha_3
\pm \frac{\epsilon^2}{2} (\alpha_+ - \alpha_- )=0.
\label{alphapmeq}
\ee
\be
(-\kappa^2+k^2 + \epsilon^2 + \Omega^2 ) \alpha_3 - \epsilon k_z (\alpha_+ -\alpha_-)=0.
\label{alpha3eq}
\ee

{\it Constraints:}
\ba
&& \hskip -1 cm
(\kappa \pm \Omega  ) k \gamma_\pm
\mp \epsilon ( \kappa \mp \Omega ) (s\beta_3+c\gamma_3)
\nn \\ && \hskip 2 cm
- \frac{2\Omega}{g} [ \kappa  +  \Omega+2\omega ] \barzeta_\pm =0
\label{gausspmeq}
\ea
\ba
&&
\kappa k \gamma_3
- \frac{\epsilon}{2} [ (\kappa + 2 \Omega ) (s\beta_+ +c\gamma_+)
\nn \\ && \hskip 2 cm
-(\kappa - 2 \Omega ) (s\beta_- +c\gamma_-)]
\nn \\ && \hskip 1.5 cm
- \frac{\Omega}{g} [ (\kappa  +  2\omega ) \barxi_+ 
- (\kappa  -  2\omega ) \barxi_- ] = 0
\label{gauss3eq}
\ea

{\it $\beta_a$ equations:}
\ba
&& \hskip -1 cm
(-\kappa^2+k^2\mp 2\Omega \kappa ) \beta_\pm \mp 2\epsilon k_z \beta_3
\pm\epsilon s k \gamma_3 \nn \\ 
&& \hskip -0. cm
\pm \frac{\epsilon^2}{2} c [ c (\beta_+-\beta_-) 
-s (\gamma_+-\gamma_-) ] 
\nn \\ && \hskip 2.5 cm
- \epsilon \frac{\Omega}{g} s (\barxi_+ + \barxi_-) = 0
\label{betapmeq}
\ea
\ba
&&\hskip 0 cm
(-\kappa^2+k^2 + c^2 \epsilon^2 + \Omega^2 ) \beta_3 
\nn \\ && \hskip -0.5 cm
+\frac{\epsilon}{2} \biggl [ s k (\gamma_+ - \gamma_-) 
- 2 c k (\beta_+ - \beta_-)
- 2 \epsilon s c \gamma_3 \biggr ] =0.
\label{beta3eq}
\ea

{\it $\gamma_a$ equations:}
\ba
&&
(-\kappa^2 \mp 2\Omega \kappa ) \gamma_\pm \pm \epsilon s k \beta_3
\nn \\ && \hskip 1 cm
\mp  \frac{\epsilon^2}{2} s [ c (\beta_+ - \beta_-) - s (\gamma_+ - \gamma_-)\} ]
\nn \\ && \hskip 2 cm
 - \epsilon \frac{\Omega}{g} c (\barxi_+ + \barxi_-) + 2 \frac{\Omega}{g}  k \barzeta_\pm = 0
 \label{gammapmeq}
 \ea
\ba
&&
(-\kappa^2+ s^2 \epsilon^2 + \Omega^2 ) \gamma_3
+\frac{\epsilon}{2} \biggl [ s k (\beta_+ - \beta_-) 
- 2 \epsilon s c\beta_3 \biggr ] 
\nn \\ && \hskip 4 cm 
+  \frac{\Omega}{g} k (\barxi_+ - \barxi_-) = 0
\label{gamma3eq}
\ea

{\it $\barxi_\pm$ equations:}
\ba
&& \hskip -1 cm
(-\kappa^2+ k^2 \mp 2\omega \kappa + 2\lambda \eta^2 )\barxi_\pm  + 2\lambda \eta^2  \barxi_\mp 
\mp \epsilon c k \barzeta_\pm 
\nn \\ && \hskip -1 cm
- \frac{\epsilon \Omega g}{4} [ (s \beta_+ +c \gamma_+) + (s \beta_- + c \gamma_-) ] 
\pm \frac{\Omega g}{2} k \gamma_3 = 0
\label{barxipmeq}
\ea

{\it $\barzeta_\pm$ equations:}
\ba
&& \hskip -1 cm
[-\kappa^2+ k^2  -\Omega (\Omega + 2\omega )  \mp 2(\omega +\Omega) \kappa ] \barzeta_\pm
\nn \\ && \hskip 2 cm
\mp \epsilon c k \barxi_\pm \pm  \frac{g \Omega}{2} k \gamma_\pm = 0
\label{barzetapmeq}
\ea

\,


\bibstyle{aps}
\bibliography{paper}

\end{document}